\documentstyle[12pt,times,dina4p]{article}

          \def\dt{\cal}
          \def\dA{{\dt A}}
          \def\dB{{\dt B}}

          \def\dF{{\dt F}}

          \def\dM{{\dt M}}

          \def\C{{\cal C}}
          \def\D{{\cal D}}

          \def\H{{\cal H}}

          \def\K{{\cal K}}
          
          \def\M{{\cal M}}
          
          \def\O{{\cal O}}
          \def\P{{\cal P}}

          \def\gD{\Delta}

          \def\gL{\Lambda}
          \def\gO{\Omega}
          
          \def\gQ{\Theta}
          \def\gS{\Sigma}

          \def\gb{\beta}

          \def\eps{\varepsilon}
          
          \def\gg{\gamma}

          \def\gs{\sigma}

          \def\Ad{{\rm{Ad}}}
          \def\aloc{\dA_{\rm loc}}
          \def\aqloc{\widetilde{\dA}}

          \def\complex{{\bf C}}

          \def\Halmos{\quad\hfill$\Box$}
          \def\id{\mbox{id}}
          \def\Jph{J_\frac{\phi}{2}}

          \def\modgrmt{\gD^{-it}}
          \def\modgrt{\gD^{it}}
          \def\modop{\gD^{\frac{1}{2}}}
          
          \def\modopt{\gD^{it}}
          \def\modopmt{\gD^{-it}}
          
          \def\oa{\O_{1}}
          
          \def\ob{\O_{2}}

          \def\pct{P$_1$CT}

          \def\Rd{\reals^{1+s}}

          \def\reals{{\bf R}}

          \def\upg{\widetilde{\P_{+}^{\uparrow}}} %universelle Ueberl.
          \def\Vpq{\overline{V}_+}

          \def\Vpq{\overline{V}_+}
          
          \def\xv{\vec x}
\title{A New Approach to Spin \& Statistics}
\author{Bernd Kuckert\thanks{
Universit\"at Hamburg, II. Institut f\"ur Theoretische
Physik, Luruper Chaussee 149, 22761 Hamburg, Germany, e-mail:
i02bku@dsyibm.desy.de}}
\date{DESY 94-211, hep-th/9412130}
\begin{document}
\frenchspacing
\maketitle

\begin{abstract}
We give an algebraic proof of the spin-statistics connection
for the para\-bo\-so\-nic and para\-fermi\-onic quantum topological
charges of a
theory of local observables with a modular \pct-symmetry.
The argument avoids the use of the spinor calculus and also
works in 1+2
dimensions.
It is expected to be a
progress towards a general spin-statistics theorem
including also (1+2)-dimensional theories with braid group statistics.
\end{abstract}

\newpage

\section{Introduction}
The spin-statistics theorem due to Fierz and Pauli \cite{Fie39,
Pau40} is one of the great
successes of general Quantum Field Theory.
A general proof in the Wightman framework
can be found in the monograph by
Streater and Wightman \cite{SW64}. However, the Bose-Fermi
alternative enters the Wightman framework via the basic
assumption of normal commutation relations.

In the algebraic approach due
to Haag and Kastler \cite{HK64, Haa92}, the Bose-Fermi alternative
is a result, not an axiom. The input of the
theory is a net $\dA$ of C$^*$-algebras $\dA(\O)$ of bounded linear
operators in a Hilbert space which are associated with every open,
bounded space-time region $\O\subset\Rd$ in such a way that
operators belonging to
spacelike separated regions commute ({\em locality}). The basic
structures of charged
fields -- including the possible particle statistics --
can be recovered from the mere observable input \cite{DHR 3,DHR 4,BF82}.
In the most general case in lower dimensions, particles
violating the familiar Bose-Fermi alternative can occur. The
spin of these particles no longer needs to be
integer or half-integer, it may be any real number.
Such particles are expected to play a role in the theory of the
fractional quantum Hall effect \cite{Sto92}.

A field net consisiting
of von Neumann algebras which generates in particular all massive
parabosonic and parafermionic sectors from the vacuum and
exhibits normal commutation relations has been
constructed by Doplicher and Roberts \cite{DR90} for
any local net of observables satisfying the standard
assumptions, and such a field is unique up to unitary equivalence.

For the algebraic framework, the
spin-statistics theorem in (1+3)-dimensional spacetime
has been proven in \cite{DHR 4} for charges which are localizable
in bounded open sets and in \cite{BE85} for charges which are
localizable in open convex
                    cones extended to spacelike infinity (spacelike
cones, see the definition below); such charges
appear in purely massive theories \cite{BF82}.
All these proofs use the spinor calculus. This structure relies on the
fact that in 1+3 dimensions
        the universal covering of the rotation group SO(3)
is of order two. In 1+2 dimensions, however, the rotation group
is the circle, and the universal covering of the circle is
of infinite order.
This is why the familiar spinor structure does not describe the
irreducible representations of $\upg$ in 1+2 dimensions
and why we have decided to look for an alternative argument.

In this article, we present
an algebraic proof
of the spin-statistics connection given in \cite{Kuc94} for
parabosonic and parafermionic charges localizable in spacelike
cones. Our proof works in any theory of
local observables in at least 1+2 dimensions with
the property that a certain antiunitary operator
$J_\dA$, a modular conjugation
associated with the net of observables and the
vacuum vector by means of the modular theory of Tomita and Takesaki,
is a \pct-operator;
here P$_1$ denotes the reflection $(x_0,x_1,x_2,\dots,x_s)
\mapsto(x_0,-x_1,x_2,\dots,x_s)$, and C
and T denote, as usual, charge conjugation and time reflection.
This assumption, which we shall refer to as {\em modular \pct-symmetry},
                has been shown by Bisognano and Wichmann
to hold for any $\upg$-covariant Wightman field \cite{BW75,BW76}.
Recently Borchers \cite{Bor92}
has proved that in 1+1 dimensions, every local net of
observables may be extended to a local net which exhibits the modular
symmetries established by Bisognano and Wichmann.
For higher dimensions, Borchers' result implies that
the modular objects considered have commutation relations
                                                with the translations
like \pct-reflections or Lorentz-boosts, respectively.
Using this fact, it can be shown
that \pct-symmetry is the only symmetry that {\em can} be implemented
on the net of observables by the considered modular conjugation
(as soon as, in a well-defined, very general sense,
{\em any} symmetry on the net of observables is implemented)
\cite{Kuc94}. On the other hand, Yngvason \cite{Yng94} has given
examples of theories which are not Lorentz-covariant and
where the modular objects considered
do not implement any symmetry.

Our line of argument is as follows: from modular \pct-symmetry
and a compactness assumption discussed below, we derive that
any rotation is represented by a product of two \pct-operators
(i.e. the \pct-operators with respect to two, in general different,
Lorentz frames).
We then transfer this result and modular \pct-symmetry
from the net of observables to the $\upg$-covariant
Bose-Fermi field constructed by
Doplicher and Roberts. The straightforward computation of
any rotation by $2\pi$ in the corresponding representation,
finally, yields the Bose-Fermi operator of the field; this
implies the familiar spin-statistics connection.

The first proof of the spin-statistics theorem
using the structures established by the Bisognano-Wichmann
theorem has been given by Fr\"ohlich and Marchetti \cite{FM92}.
Their argument relies on the assumption of the {\em full}
Bisognano-Wichmann structure
not only for the net of local observables,
but for the whole reduced field bundle (which does not consist
of algebras). We here only make an
assumption about the net of observables, namely
modular \pct-symmetry; we need not assume anything concerning
the Bisognano-Wichmann modular operator or modular group.

A proof of the spin-statistics theorem
                               similar to ours has been found
independently by Guido and Longo \cite{Lon94}.
They have derived
modular \pct-symmetry from the assumption that a certain
modular group implements a one-parameter group of
Lorentz boosts ("modular covariance").
\section{Notation, Preliminaries, and Assumptions}\label{notation}

For some integer $s\geq2$,
denote by $\Rd$ the (1+s)-dimensional Minkowski space, and let $V_+$
be the forward light cone. The set $\K$ of all {\em double cones},
i.e. the set of all open sets $\O$ of the form
$$\O:=(a+V_+)\cap(b-V_+),\qquad a,b\in\Rd,$$
is a convenient topological base of $\Rd$. Each nonempty double
cone is fixed by two points, its upper and its lower apex, and
the set $\K$ is invariant under the action of the Poincar\'e group.

In the sequel, we denote by $(\H_0,\dA)$ a {\em concrete local
net of observables}: $\H_0$ is a Hilbert space, and the net $\dA$
associates with every double cone $\O\in\K$
a (concrete) C$^*$-algebra $\dA(\O)$ consisting
of bounded operators in $\H_0$
and containing the identity operator; this
mapping is assumed to be {\em isotonous}, i.e. if $\oa\subset\ob$,
                                     $\oa,\ob\in\K$,
then $\dA(\oa)\subset\dA(\ob)$, and {\em local}, i.e.
if $\oa$ and $\ob$ are spacelike separated double cones and
if $A\in\dA(\oa)$, $B\in\dA(\ob)$, then $AB=BA$.
Since $\K$ is a topological base of $\Rd$, we may for any open set
$M\subset\Rd$ consistently define $\dA(M)$ to be the C$^*$-algebra
generated by the C$^*$-algebras $\dA(\O)$, $\O\in\K$, $\O\subset M$.
We shall denote by $\aqloc:=\dA(\Rd)$ the C$^*$-algebra of {\em
quasilocal observables}. Note
that every state of the normed, involutive algebra
$\aloc=\bigcup_{\O\in\K}\dA(\O)$ of all {\em local observables}
has a continuous extension to a state of $\aqloc$.

For any subset $M$ of $\Rd$, we denote by $M'$ the
{\em spacelike complement} of $M$, i.e. the set of all points in
$\Rd$ which are spacelike with respect to all points of $M$,
and for every algebra $\M$ of bounded operators in some Hilbert space
$\H$, we denote by $\M'$ the algebra of all bounded operators
which commute with all elements of $\M$.
Using this notation, the above locality assumption  reads
$\dA(\O)\subset\dA(\O')'\quad\forall\O\in\K$.

Another kind of regions in Minkowski space that will be used
are the {\em spacelike cones}:
for any open, salient,
convex circular cone $\vec\C$ in $\reals^s$, i.e. for any cone in
       $\reals^s$
which is generated by some open $\eps$-ball around a vector $\xv\in
\reals^s$ with euclidean length $\|\xv\|_2>\eps$, the
causal completion $\C$ of $\vec\C$ and its Poincar\'e
transforms will be called spacelike cones; their set will be
denoted by $\gS$. Note that
this definition, which is based on remarks in \cite{BE85},
singles out the {\em causally complete} spacelike cones in the
sense
of \cite{BF82}, i.e. cones with $\C''=\C$.

\subsection*{Assumptions, I:}
\begin{quote}{\it
We shall assume that there exists in $\H_0$
a strongly continuous representation $U$
of the universal covering $\upg$ of the
restricted Poincar\'e group $\P_+^\uparrow$ and that $\dA$ is
{\em covariant} with respect to $U$, i.e.
$$U(g)\dA(\O)U(g)^*=\dA(\gL(g)\O)\qquad\forall g\in\upg,$$
where $\gL:\,\upg\to\P^\uparrow_+$ denotes the covering map.

We assume that the translations in
               $U$ satisfy the {\em spectrum condition}, i.e.
the joint spectrum of their generators is contained in $\Vpq$.

We assume the existence and uniqueness up to a phase of a
unit vector $\gO$ in $\H_0$ which is invariant under $U$ and
{\em cyclic} with respect to the concrete algebra $(\H_0,\aqloc)$,
i.e. $\overline{\aqloc\gO}=\H_0$;
$\gO$ will be called the {\em vacuum vector}.

The (vacuum) representation $(\H_0,\id_{\aqloc})$ is
assumed to be irreducible, i.e.}
$$\aqloc'=\complex\,\id_{\H_0},$$
{\it and we assume $\H_0$ to be separable.}
\end{quote}
Buchholz and Fredenhagen have associated a unique irreducible
vacuum representation $(\H_{\rm vac},\pi_{\rm vac})$ with any massive
                           single-particle
representation $(\H,\pi)$ of $\aqloc$ (\cite{BF82}, Definition on
p. 13 and Theorem 3.4). If $(\H,\pi)$ is irreducible,
it is unitarily equivalent to $(\H_{\rm vac},\pi_{\rm vac})$
when restricted to $\dA(\C')$ for any spacelike cone $\C$
                                                     (\cite{BF82},
Theorem 3.5), so any irreducible massive single-particle
representation may be regarded as
an excitation of a vacuum. We may choose such a vacuum fixed for our
purposes; therefore, we set $(\H_{\rm vac},\pi_{\rm
vac})=(\H_0,\id_{\aqloc})$. We shall denote the set of
all parabosonic and parafermionic spacelike-cone excitations
(in the above sense) of the
vacuum by $\Pi_\gS$. We need not confine
ourselves to massive particle representations, however,
we do not know any examples of representations which are localizable
in spacelike cones and have finite statistics but do not
arise from massive single-particle representations.

We shall repeatedly use the group homomorphism $r:\,\reals\to\upg$
which is constructed as follows: denote by $\exp(i\cdot)$ the
covering map $\phi\mapsto\exp(i\phi)$ from $\reals$ onto
$S^1$, and let $\iota:\,S^1\to\P_+^{\uparrow}$ be the group homomorphism
embedding
$S^1$ into $\P_+^{\uparrow}$ as the group of rotations in the
1-2-plane. $\exp(i\cdot)$ and $\iota$ are continuous,
so $\iota\circ\exp(i\cdot)$ is a continuous curve in
$\P_+^{\uparrow}$. There is a unique lift of this curve to
a continuous curve $r$ in $\upg$ with $r(0)=1_{\upg}$
(see, e.g., Theorem III.3.3. in \cite{Bre93});
$r$ is a group homomorphism of $\reals$ into $\upg$.

Denote now by $W$ the wedge
$$W:=\{x\in\Rd:\,x_1\geq|x_0|\}.$$ From
Theorem 1 in \cite{Bor68} (cf. also
\cite{Buc74}, p. 279) it
follows as a modification of the Reeh-Schlieder theorem
that $\gO$ is cyclic with respect to
$(\H_0,\dA(W))$, so a fortiori with respect to $(\H_0,\dA(W)'')$,
and using a standard argument (see, e.g., Prop. 2.5.3 in \cite{BraRo}),
one obtains from locality that $\gO$
is also {\em separating} with respect to $(\H_0,\dA(W)'')$, i.e.
if $A\in\dA(W)''$ and $A\gO=0$, then $A=0$.

A triple $(\H,\dM,\xi)$ consisting of a von Neumann algebra $(\H,\dM)$
and a cyclic and separating vector $\xi$ is called a {\em standard von
Neumann algebra}; such a triple is the setting
of Tomita-Takesaki theory (\cite{Tak70}, see also \cite{BraRo,Haa92}):
                                       the antilinear operator
$$S_0:\,\dM\xi\to\dM\xi;\,
  A\xi\mapsto A^*\xi,\quad A\in\dM,$$
is closable, and its closure $S$
                             admits a (unique) polar decomposition
into
an antiunitary operator $J$ and a positive
operator $\modop$ defined on the domain of $S$ such that
$$S=J\modop.$$
$J$ is called the {\em modular conjugation}, $\gD=(\modop)^2$ the
{\em modular operator} of the standard von Neumann algebra
$(\H,\dM,\xi)$. The unitary group $(\modopt)_{t\in\reals}$
is called the {\em modular group} of $(\H,\dM,\xi)$.
$\gD$, $\modopt$, $t\in\reals$, and $J$, which we refer to as the
                     {\em modular objects} of
the standard von Neumann algebra $(\H,\dM,\xi)$, leave $\xi$
invariant. From the well-known elementary
relations of the modular objects (Theorem 7.1. in \cite{Tak70}) we shall
use $J^2=\id_{\H}$.
We recall the theorem of Tomita and Takesaki \cite{Tak70}:
\begin{eqnarray*}
\modopt\dM\modopmt&=&\dM;\\
J\dM J&=&\dM'.
\end{eqnarray*}
We shall make use of two further basic facts from Tomita-Takesaki
theory; we recall them for the reader's convenience:
\subsection{Lemma} \label{Lemma A}
\begin{quote}
{\it Let $(\H_1,\dM_1,\xi_1)$ and $(\H_2,\dM_2,\xi_2)$
be two standard von Neumann algebras with modular objects $\gD_1$,
$J_1$ and $\gD_2$, $J_2$, respectively, and let $V: \H_1\to\H_2$
be a unitary operator
             with $V\dM_1 V^*=\dM_2$ and $V\xi_1=\xi_2$. Then we have
$V\modop_1 V^*=\modop_2$ and
$VJ_1 V^*=J_2$.}
\end{quote}

\subsection{Theorem (Takesaki, Winnink)}\label{KMS}
\begin{quote}{\it
For any standard von Neumann algebra $(\H,\dM,\xi)$, the
modular automorphism group $\left(\Ad(\modopt)\right)_{t\in\reals}$
is the
unique one-parameter group $(\gs_t)_{t\in\reals}$ of
automorphisms of the von Neumann algebra $(\H,\dM)$ which
satisfies the following conditions:

(i) for any $A\in\dM$, the function
     $t\mapsto\gs_t(A)$, $t\in\reals$,
                        is a continuous function from
$\reals$ into the von Neumann algebra $(\H,\dM)$ endowed with
the strong operator topology;

(ii) $(\gs_t)_{t\in\reals}$
               satisfies the {\em KMS-condition (at the inverse
temperature $\gb=1$) with respect to} $(\H,\dM,\xi)$:
      for any $A,B\in\dM$, the function
$t\mapsto\langle\xi,A\gs_t(B)\xi\rangle$, $t\in\reals,$ may be
extended to a continuous function $f$ on the complex strip
$0\leq{\rm Im}\,z\leq 1$ which is analytic on the interior
of this strip and satisfies}
$$f(t+i)=\langle\xi,\gs_t(B)A\xi\rangle\qquad\forall t\in\reals.$$
\end{quote}
The proof of Lemma \ref{Lemma A} is straightforward; for a proof
of Theorem \ref{KMS}, see \cite{Tak70}, Theorems 13.1 and 13.2; note
that
(ii) implies that $\langle\xi,\gs_t(A)\xi\rangle=\langle\xi,A\xi\rangle
\,\forall A\in\dM$, $t\in\reals$
                       since $t\in\reals
\mapsto\langle\xi,\gs_t(A)\xi\rangle$ is a bounded function for any
$A\in\dM$.

As mentioned in the Introduction, Bisognano and Wichmann have
             shown that
the modular objects $\modop_\dA$, $\modopt_\dA$, $t\in\reals$,
and $J_\dA$ of the standard von Neumann algebra $(\H_0,\dA(W)'',\gO)$
implement symmetries in any Wightman theory;
in particular, $J_\dA$ implements a \pct-symmetry.
There is, at the moment, no reason to believe that this form of
\pct-symmetry should only hold for Wightman fields.

\subsection*{Assumption II:}
\begin{quote}{\it
Denote by $j$ the P$_1$T-reflection given by
$$j(x_0,x_1,x_2,\dots x_s):=(-x_0,-x_1,x_2,\dots,x_s);$$
we shall assume {\em modular \pct-symmetry}:
$$J_\dA\dA(\O)J_\dA=\dA(j\O)\qquad\forall\O\in\K.$$
}\end{quote}
Note that due to $\upg$-covariance, this condition automatically
holds in all Lorentz frames as soon as it holds in one.
Given modular \pct-symmetry,
$J_\dA$ indeed yields the correct charge
conjugation \cite{GL92}. In 1+3 dimensions, a full PCT-operator may be
constructed as a product of such modular conjugations; in 1+2
dimensions, no full PCT-symmetry has ever been proved to exist
(cf. also \cite{Mon69} for a discussion in the Wightman framework).

 From the Tomita-Takesaki theorem it follows that
modular \pct-symmetry implies {\em wedge duality}:
$$\dA(W)''=\dA(W')'.$$
This, again,
     implies the following duality assumption for spacelike cones:
$$\dA(\C')'=\dA(\C)''\qquad\forall\C\in\gS$$
since for any two spacelike separated spacelike cones $\C_1$ and $\C_2$,
there is a Poincar\'e transform $\hat W$ of $W$ such that $\C_1\subset
\hat W$ and $\C_2\subset\hat W'.$

The adjoint action $\Ad(j)$
                   of $j$ on $\P_+^{\uparrow}$ has a unique lift to
a group homomorphism of $\upg$ (cf. Section III.4 in \cite{Bre93})
                               which we shall denote by
$\widetilde{\Ad}(j)$.

\subsection*{Assumption III:}
\begin{quote}{\it
The group of {\em internal symmetries} of $(\H_0,\dA,\gO)$,
i.e. the group of all unitaries $\gg$ in $\H_0$ such that
$\gg\gO=\gO$ and
$\gg\dA(\O)\gg^*=\dA(\O)$ for all $\O\in\K$,
is assumed to be compact in the strong operator topology.
}\end{quote}
This property has been derived in \cite{DHR 1} from assumptions
concerning the scattering theory of the system.
Another sufficient condition
is the distal split property \cite{DL84}.
The distal split property, again,
has been derived by Buchholz and Wichmann from
their so-called {\em nuclearity condition}, for which they
have given a thermodynamical justification \cite{BuW86}.

On the other hand,
the compactness of the internal symmetries implies that all
internal symmetries commute with all
$U(g)$, $g\in\upg$, and that $U$ is the unique strongly continuous
unitary representation
of $\upg$ in $\H_0$ with respect to which $\dA$ is covariant
and $\gO$ is invariant \cite{DL84,BGL93}. There are
examples of $\upg$-covariant theories \cite{Str67}
which violate the familiar
spin-statistics connection. They admit several unitary representations
of $\upg$ under which they are covariant, so, a fortiori, our
compactness assumption is violated.

Finally, we recall the definitions and results of the Doplicher-Roberts
field construction performed in \cite{DR90} which are used in our
argument.
\subsection{Definition} \label{field}
\begin{quote}
Let $(\H_0,\dA,U,\gO)$ be as above, let $\H$ be a (not necessarily
separable) Hilbert space,
and let $(\dF(\C))_{\C\in\gS}$ be a net\footnote{Here we make an
abuse of language: the index set $\gS$ is not directed, so
$(\dF(\C))_{\C\in\gS}$ is not a net in the usual sense. In this
paper, we also call a net any family (of algebras) indexed
by a partial-ordered, not necessarily directed
set (of regions in Minkowski space).}
of von Neumann
algebras. Let $\pi$ be a faithful representation of $\aqloc$ in
$\H$, and let $G$ be a strongly compact group of unitaries in $\H$.
The quadruple $(\H,\dF,\pi,G)$ is called an {\em extended field
system with gauge symmetry} --- we shall simply say: a {\em field} ---
{\em over} $(\H_0,\dA,U,\gO)$ if the following conditions are satisfied:

(i) $(\H,\pi)$ contains $(\H_0,\id_{\aqloc})$ as a subrepresentation;

(ii) $\H_0$ is the subspace of all $G$-invariant vectors in $\H$;

(iii) for any $\C\in\gS$, the maps $\Ad(\gg)$, $\gg\in G$, act as
automorphisms on $\dF(\C)$, and $\pi(\dA(\C))''$ is the algebra
of those elements of $\dF(\C)$ which are invariant under all
$\Ad(\gg)$, $\gg\in G$, i.e.:
$$\pi(\dA(\C))''=\dF(\C)\cap G'\qquad\forall\C\in\gS;$$

(iv) $\dF$ is {\em irreducible} and {\em weakly additive}:
$$\left(\bigcup_{a\in\Rd}\dF(\C+a)\right)''=\dB(\H)\qquad\forall
\C\in\gS;$$

(v) $\dF$ has the {\em Reeh-Schlieder property for spacelike
cones}:
$$\overline{\dF(\C)\gO}=\H\qquad\forall\C\in\gS;$$

(vi) $\dF$ is {\em local with respect to the net}
$(\pi(\dA(\O)))_{\O\in\K}$:
$$\dF(\C)\subset\pi(\dA(\C'))'\qquad\forall\C\in\gS.$$

In this case, $\dF$ is called the {\em field net}, and $G$ is called
the {\em (global) gauge group}.

A field $(\H,\dF,\pi,G)$ is called {\em normal} if it satisfies the
{\em normal commutation relations}, i.e., if the gauge group
contains an involution $k$ such that with the notations
$$F^\pm:=\mbox{$\frac{1}{2}$}(F\pm kFk^*),\qquad F\in\dF(\C),
\C\in\gS,$$
we have for any two spacelike separated cones $\C_1$ and $\C_2$:
$$F_1^+F_2^+=F_2^+F_1^+,\qquad F_1^+F_2^-=F_2^-F_1^+,\qquad
F_1^-F_1^-=-F_2^-F_1^-\qquad\forall F_{1,2}\in\dF(\C_{1,2}).$$
$k$ is called a {\em Bose-Fermi operator}.
\end{quote}
Using the separability of $\H_0$, Doplicher and Roberts have shown
that, given any field $(\H,\dF,\pi,G)$ over $(\H_0,\dA,U,\gO)$,
every irreducible subrepresentation of $(\H,\pi)$ is
contained in $\Pi_\gS$ (Theorem 3.6. in \cite{DR90},
cf. also the remarks on p. 19 in \cite{BF82}), i.e., for some index set
$I$, there is a family $(\pi_\iota)_{\iota\in I}$ of irreducible
representations in $\Pi_\gS$ such that $\pi=\bigoplus_{\iota\in I}
\pi_\iota$. If the field is normal and $k$ is a Bose-Fermi operator
of the field, then for every
$\iota\in I$, the restriction of the Bose-Fermi
operator $k$ to $\H_\iota$ coincides with the sign of the
                                              statistics parameter
of $\pi_\iota$ (Theorem 3.6. in \cite{DR90}). Hence, $k$ is
uniquely determined by $\pi$.

If $(\H,\dF,\pi,G)$ is a normal field over $(\H_0,\dA,U,\gO)$,
the unitary operator defined by
$$V:=\frac{1}{1+i}(\id_{\H}+ik)$$
implements a {\em twist} of the field: the field over $(\H_0,\dA,U,\gO)$
given by
$$\dF^t(\C):=V\dF(\C)V^*,\qquad\C\in\gS,$$
is local with respect to $\dF$, i.e. $\dF(\C)\subset\dF^t(\C')'$
for all $\C\in\gS$.
Doplicher and Roberts even established {\em twisted duality}, i.e.
$$\dF(\C)=\dF^t(\C')'\qquad\forall\C\in\gS.$$
See Theorem 5.4. in \cite{DR90}.
The same arguments allow to show that the wedge duality of the net of
observables implies {\em twisted wedge duality} of the field:
$$\dF(W)=\dF^t(W')',$$
               where
$$\dF(W):=\left(\bigcup_{{\C\in\gS}\atop{\C\subset W}}
                 \dF(\C)\right)''.$$
Note that the phase $\frac{1}{1+i}$ of $V$
has been chosen such that $V$ leaves $\gO$ invariant;
with this choice, $V^2=k$.

Given, conversely, $(\H_0,\dA,U,\gO)$ as assumed above,
Doplicher and Roberts have shown that
there is an up to unitary equivalence unique normal
                                             field $(\H,\dF,\pi,G)$ over
$(\H_0,\dA,U,\gO)$ such that each irreducible representation in
$\Pi_\gS$ is unitarily equivalent
to a subrepresentation of $(\H,\pi)$
(Theorem 5.3 in \cite{DR90}\footnote
{
Note that the full $\upg$-covariance is not needed at this
stage; it would suffice to assume translation covariance.

Furthermore we remark that
Doplicher and Roberts make an additional
assumption they call "property B'".
However, the following is sufficient:

\smallskip\noindent%
{\bf Borchers property for spacelike cones:}
\begin{quote}
A concrete net $(\H_0,\dA)$ of observables is said to have the
{\em Borchers property for spacelike cones} if, given any two spacelike
cones $\C_1$ and
$\C_2$ with $\overline{\C_1}\subset\C_2$ which are chosen
in such a way that there is a third spacelike cone
$\C^{\times}$ with $\C^{\times}\subset\C_1'\cap\C_2$,
we can find for each nonzero projection $E\in\dA(\C_1)''$ an isometry
$W\in\dA(\C_2)''$ such that $WW^*=E$ (and, trivially, $W^*W=\id_{\H_0}$,
i.e., $E$ and $\id_{\H_0}$ are equivalent in $\dA(\C_2)''$).
\end{quote}
Noting that for any spacelike cone $\C$, we have additivity:
$$\left(\bigcup_{a\in\Rd}\dA(\C+a)\right)''=\aqloc''=\dB(\H_0),$$
and using the spectrum condition and irreducibility, the Borchers
                                property for spacelike
cones can be
proven applying the arguments from \cite{Bor67}. We emphasize that
Borchers proves in \cite{Bor67}
                the corresponding result for double cones and
therefore has to {\em assume} for double cones the above
additivity property.

Doplicher's and Roberts' property B' is stronger: the same
assumption as in the Borchers property for spacelike cones
is made for any two spacelike cones $\C_1$ and $\C_2$ with
$\overline{\C_1}\subset\C_2$ even if there is no
spacelike cone $\C^\times\subset\C_1'\cap\C_2$ (this is, e.g., the case
if $\C_1$ is a translate of $\C_2$). However, in order
to prove this stronger
form of the Borchers property for spacelike cones by means of
the arguments taken from \cite{Bor67}, one has to assume weak
additivity for double cones.
}.
There also is, up to unitary equivalence, a unique normal field
$(\H,\dF,\pi,G)$ over $(\H_0,\dA,U,\gO)$
such that $(\H,\pi)$ contains all irreducible
{\em $\upg$-covariant} representations contained in the set $\Pi_\gS$
and
is, conversely, a direct sum of such representations
                               (\cite{DR90}, top of p. 98).
There is a unique strongly continuous unitary representation $U_\pi$
of $\upg$ in $\H$ with
$$U_\pi(g)\pi(A)U_\pi(g)^*=\pi(U(g)AU(g)^*)\qquad\forall g\in\upg,\,A\in
\aqloc$$
(\cite{DR90}, pp. 98-101, cf. also Lemma 2.2. in \cite{DHR 4}).
The vacuum vector is invariant under $U_\pi$, and the field net
$\dF$ is covariant with respect to $U_\pi$. Note that $U_\pi$
does not depend on the field net $\dF$ itself\footnote
{As an example, consider some $\upg$-covariant field $(\H,\dF,\pi,G)$
and its twisted field $(\H,\dF^t,\pi,G)$. Both are covariant under
$U_\pi$ although their field nets are different.}.
Such a field will be called a {\em $\upg$-covariant} (normal) field
over $(\H_0,\dA,U,\gO)$.

It follows from property (v) in Definition \ref{field} that $\gO$
is cyclic and separating with respect to the von Neumann algebra
$(\H,\dF(W))$.
We shall denote by $J_\dF$ and $\gD_\dF$ the modular conjugation
and operator of the standard von Neumann algebra $(\H,\dF(W),\gO)$.
\section{Results}

\subsection{Proposition}\label{mod com rel}
\begin{quote}{\it
For any $g\in\upg$, we have}
$$J_\dA U(g)J_\dA=U(\widetilde{\Ad}(j)g).$$
\end{quote}

{\bf Proof:} One easily verifies that the representation $U$
and the strongly continuous unitary representation $U^J$ of $\upg$
                                                         defined by
$$U^J(g):=J_\dA U(\widetilde{Ad}(j)g)J_\dA,\qquad g\in\upg,$$
implement the same spacetime transformations on the net
$\dA$ and leave $\gO$ invariant. As stated in the previous
section, it follows from the strong compactness of the
group of internal symmetries that there can be at most one
such representation; this implies $U=U^J$.
\Halmos

\subsection{Corollary (modular rotation symmetry)}\label{mod rot sym}
\begin{quote}{\it
For every angle $\phi\in[0,2\pi]$, denote
       by $W_\phi$ the rotation by $\phi$ of $W$ in the 1-2-plane
and by $J_\phi$ the modular conjugation of
$(\H_0,\dA(W_\phi)'',\gO)$. With $r$ as defined in the previous
section, define
$R(\phi):=U(r(\phi))$, $\phi\in\reals$.
We then have
$$R(\phi)=\Jph J_\dA\qquad\forall\phi\in\reals.$$
The representation $U$ does not only realize $\upg$-covariance, but
even (restricted) Poincar\'e covariance of the net:
$$R(2\pi)=\id_{\H_0}.$$
}\end{quote}
{\bf Proof:} From Proposition \ref{mod com rel}, we get
$$J_{\frac{\phi}{2}}J_\dA=R(\mbox{$\frac{\phi}{2}$})J_\dA
R(\mbox{$-\frac{\phi}{2}$})
J_\dA=R(\mbox{$\frac{\phi}{2}$})R(\mbox{$\frac{\phi}{2}$})J_\dA^2
=R(\phi).$$
In particular,
$$R(2\pi)=J_\pi J_\dA=J_\dA^2=\id_{\H_0};$$
in the second step we use that the modular conjugations of
a standard von Neumann algebra and its commutant coincide.
\Halmos

\bigskip\noindent%
For the special case of the rotation by $\pi$ in (1+1)-dimensional
chiral theories, the above
formula has already appeared in \cite{Sch92,Wie93}.

\subsection{Lemma} \label{ext mod obj}
\begin{quote}
{\it
For any field $(\H,\dF,\pi,G)$ over $(\H_0,\dA,U,\gO)$, we have
$$\begin{array}{lrrll}
\makebox{(i)}
&\modgrt_\dF|_{\H_0}A\modgrmt_\dF|_{\H_0}&=&\modgrt_\dA
                                A\modgrmt_\dA
&\forall A\in\dA(W)'';\\
\makebox{(ii)}
&\modgrt_\dF|_{\H_0}&=&\modgrt_\dA;&\\
\makebox{(iii)}
&\modop_\dF|_{\H_0\cap D(\modop_\dF)}&=&\modop_\dA;&\\
\makebox{(iv)}
&J_\dF|_{\H_0}&=&J_\dA;&\\
\makebox{(v)}
&J_\dF\pi(A)J_\dF&=&\pi(J_\dA AJ_\dA)&\forall A\in\aqloc.
\end{array}$$
}
\end{quote}
{\bf Proof:}
It follows from Lemma \ref{Lemma A} that $\modopt_\dF$ commutes with
the elements of the gauge group $G$ for any $t\in\reals$. This implies
that any such $\modopt_\dF$ maps the $G$-invariant vectors in $\H$
into $G$-invariant vectors, i.e., $\modopt_\dF\H_0=\H_0$
because of property (ii) in Definition \ref{field}, and that its
adjoint action acts as an automorphism
on the commutant of $G$. This -- together with the Tomita-Takesaki
theorem and the identity $\dF(W)\cap G'=\pi(\dA(W))''$ following
from property (iii) in Definition \ref{field} -- gives that
$\Ad(\modopt_\dF)$ acts as an automorphism on
$\pi(\dA(W))''$.

Consider now the direct sum decomposition $\pi=\bigoplus_{\iota\in
I}\pi_\iota$ of $\pi$ into irreducible representations $\pi_\iota$
in $\Pi_\gS$. The restriction of
each $\pi_\iota$ to $\dA(W)$ has a faithful extension $\pi_\iota^W$
                                               to $\dA(W)''$
which is continuous with respect to all familiar operator topologies
(cf. Lemma 4.1. in \cite{BF82}).
Noting that
$$\bigoplus_{\iota\in I}\pi_\iota^W(\dA(W)'')=\bigoplus_{\iota\in I}
(\pi_\iota^W(\dA(W))'')=\left(\bigoplus_{\iota\in I}\pi_\iota^W(\dA(W))
\right)''=(\pi(\dA(W)))''$$
(cf. Cor. II.3.6. in \cite{Tak79} for the second step), we obtain a
                                 faithful extension
$\pi_W$: $\dA(W)''\to\pi(\dA(W))''$ of $\pi$ by setting
$$\pi_W(A):=\bigoplus_{\iota\in I}\pi_\iota^W(A),\qquad A\in\dA(W)''.$$
Because of property (i) in
Definition \ref{field}, the inverse is given by
$$\pi_W^{-1}(B)=B|_{\H_0}\qquad\forall B\in\pi_W(A(W))''.$$
It is obviously continuous with respect to the corresponding
strong operator topologies (the same holds for the
other familiar operator topologies).

We may now define a one-parameter group $(\gs_t)_{t\in\reals}$
of automorphisms of $(\H_0,\dA(W)'')$ by
$$\gs_t(A):=
\pi_W^{-1}\left(\modopt_\dF\pi_W(A)\modopmt_\dF\right),
\qquad A\in\dA(W)'',$$
and since we have shown above that
          the $\modopt_\dF$, $t\in\reals$, leave the subspace $\H_0$
invariant, we conclude
$$\gs_t(A)=\modopt_\dF|_{\H_0}A\modopmt_\dF|_{\H_0}\qquad\forall
                     A\in\dA(W)'', t\in\reals.$$
 From this and from Theorem \ref{ext mod obj} it follows that $\gs$
satisfies the conditions of Theorem \ref{ext mod obj}
and therefore coincides with the modular automorphism
group of $(\H_0,\dA(W)'',\gO)$; this proves (i).

(ii) follows from (i): for any $A\in\dA(W)''$ and any $t\in\reals$,
we have
$$\modopt_\dF|_{\H_0}A\gO=\modopt_\dF|_{\H_0}A\modopmt_\dF|_{\H_0}\gO
=\modopt_\dA A\modopmt_\dA\gO=\modopt_\dA A\gO,$$
so $\modopt_\dF|_{\H_0}$ and $\modopt_\dA$ coincide on a dense
subspace of $\H_0$ and hence -- being bounded -- on all of $\H_0$.

(iii) follows from (ii) since the KMS-condition implies, in the
sense of quadratic forms:
$$\langle A\gO,\gD_\dF|_{D(\gD_\dF)\cap\H_0}B\gO\rangle
=\langle B^*\gO,A^*\gO\rangle=\langle A\gO,\gD_\dA B\gO\rangle
\qquad\forall A,B\in\dA(W)''.$$
Since any positive operator is uniquely determined by its
quadratic form,
we conclude $\gD_\dF|_{\D(\gD_\dF)\cap\H_0}=\gD_\dA$, and using the
spectral theorem, we get (iii).

(iv) follows from (iii) since the range of $\modop_\dA$ is
dense in $\H_0$ and
$$\left(J_\dF\modop_\dF\right)|_{D(\modop_\dF)\cap\H_0}
=J_\dA\modop_\dA.$$

(v) follows from (iv): because of modular \pct-symmetry, we have
$J_\dA\aqloc J_\dA=\aqloc$, hence, twice using property (i) in
Definition \ref{field}, we get
\begin{eqnarray*}
J_\dF\pi(A)J_\dF\gO&=&J_\dF\pi(A)\gO=J_\dF|_{\H_0}\pi(A)|_{\H_0}\gO\\
&=&J_\dA A\gO=J_\dA AJ_\dA\gO=\pi(J_\dA AJ_\dA)|_{\H_0}\gO\\
&=&\pi(J_\dA AJ_\dA)\gO,
\end{eqnarray*}
and since $\gO$ is separating with respect to $(\H,\dF(W)')=
(\H,\dF^t(W'))$,
statement (v) follows from the Tomita-Takesaki theorem.
\Halmos

\bigskip\bigskip\noindent%
{\bf Remark:} In the above argument, the modular groups considered
possibly do not implement any symmetry on the net $\dA$. We mention
                                      that under the
additional assumption that $\modopt_\dA\aqloc\modopmt_\dA
\subset\aqloc$, $t\in\reals$, one can derive
$$\modopt_\dF\pi(A)\modopmt_\dF=\pi(\modopt_\dA A\modopmt_\dA)
\qquad\forall A\in\aqloc, t\in\reals$$
from (ii) in the same way as we have obtained (v) from (iv)
in the preceding proof.

\subsection{Theorem (\pct-symmetry of the field)}
\begin{quote}
{\it
Let $(\H,\dF,\pi,G)$ be a $\upg$-covariant, normal field
over $(\H_0,\dA,U,\gO)$. Then we have:

(i) $J_\dF\dF(\C)J_\dF=\dF^t(j\C)\qquad\forall\C\in\gS;$

(ii) $J_\dF U_\pi(g)J_\dF=
     U_\pi(\widetilde{\Ad}(j)(g))\qquad\forall g\in\upg.$

The antiunitary involution $\gQ_W:=VJ_\dF=J_\dF V^*$ is the
\pct-operator, i.e.}
$$\gQ_W\dF(\C)\gQ_W=\dF(j\C).$$
\end{quote}
{\bf Proof:} Note first that $VJ_\dF=J_\dF V^*$ follows from the
definition of $V$ by a straightforward computation. From the modular
\pct-symmetry of $\dA$, the preceding lemma,
and the fact that the modular objects commute with internal
symmetries, it follows that by
\begin{eqnarray*}
\H^J&:=&J_\dF\H=\H;\\
\dF^{J}(\C)&:=&J_\dF\dF(j\C)J_\dF,\qquad\C\in\gS;\\
\pi^{J}(A)&:=&J_\dF\pi(J_\dA AJ_\dA)J_\dF=\pi(A),
                                         \qquad A\in\aqloc,\\
G^J&:=&J_\dF GJ_\dF=G
\end{eqnarray*}
a second $\upg$-covariant normal field $(\H^J\!=\!\H,\dF^J,\pi^J\!=\!
\pi,G^J\!=\!G)$ over $(\H_0,\dA,U,\gO)$ is defined; note that
it follows from $\pi^J=\pi$ that $\dF^J$ has the same Bose-Fermi
operator as $\dF$ and that $\dF^J$ is covariant under $U_\pi=U_{\pi^J}$.
To show that $\dF^J=\dF^t$, let $\C_1$
and $\C_2$ be two spacelike cones with
$\C_1\subset W$ and $\C_2\subset W'$. Using
the Tomita-Takesaki theorem, we get
\begin{eqnarray*}
\dF^J(\C_1)&=&J_\dF\dF(j\C_1)J_\dF\subset J_\dF\dF(W')J_\dF\\
           &=&J_\dF V^*V\dF(W')V^*VJ_\dF=VJ_\dF\dF^t(W')J_\dF V^*
            =VJ_\dF\dF(W)'J_\dF V^*\\
           &=&V\dF(W)V^*=\dF^t(W)=\dF(W')'\\
           &\subset&\dF(\C_2)'.
\end{eqnarray*}
Since for any spacelike separated cones $\C_1$ and $\C_2$,
we can find a Poincar\'e transform $\hat W$ of $W$ such that
$\C_1\subset\hat W$ and $\C_2\subset\hat W'$, the net $\dF^J$ is
easily shown to be
local with respect to the net $\dF$.
Twisted duality implies $\dF^J\subset\dF^t$, hence $\gQ_W\dF(\C)\gQ_W
\subset\dF(j\C)\,\forall\C\in\gS$. Since $\gQ_W$ is an involution,
we conclude $\gQ_W\dF(\C)\gQ_W=\dF(j\C)$ and $\dF^J=\dF^t$.
 From this, the Theorem follows immediately.
\Halmos

\subsection{Corollary (spin-statistics theorem)}
\begin{quote}{\it
Let $(\H,\dF,\pi,G)$ be a covariant, normal field over
$(\H_0,\dA,U,\gO)$.

For every angle $\phi\in[0,2\pi]$, denote by $W_\phi$ the rotation of
$W$ by $\phi$ in the 1-2-plane, and let $J_\phi$
 and $\gQ_{W_\phi}$ be the
modular conjugation and the corresponding \pct-operator of
$(\H,\dF(W_\phi),\gO)$. With $r$ as defined in the previous section,
define $R_\pi(\phi):=U_\pi(r(\phi))$, $\phi\in\reals$.

Then we have:
$$R_\pi(\phi)=\Jph J_\dF=\gQ_{W_\frac{\phi}{2}}\gQ_W.$$
In particular, $R_\pi(2\pi)=k$, i.e. the spin-statistics connection
familiar from 1+3 dimensions holds.}
\end{quote}
{\bf Proof:} The first statement immediately follows from the
preceding theorem in the same way as Corollary \ref{mod rot sym}
follows from Proposition \ref{mod com rel}.

To obtain the spin-statistics connection, note that
$J_\pi=VJ_\dF V^*$ follows from wedge duality by Lemma \ref{Lemma A}.
Since $VJ_\dF=J_\dF V^*$, we obtain
$$R_\pi(2\pi)=J_\pi J_\dF=VJ_\dF V^*J_\dF=V^2J_\dF^2=V^2=k.$$
\Halmos

\section*{Acknowledgements}
I would like to thank K. Fredenhagen for
numerous discussions.
His hints and comments have considerably accelerated and faciliated
this work. Furthermore, I would like to thank C. Adler, D. Arlt,
D. Buchholz, K. Fredenhagen, M. J\"or\ss, W. Kunhardt, N. P. Landsman
                                      and
K.-H. Rehren for their critical reading of the manuscript.

\end{document}